\documentclass[12pt]{iopart}

\begin{document}

\title{Doubly-Charged Bileptons at the LHC}

\author{Paul H. Frampton}

\address{\it Dipartimento di Matematica e Fisica "Ennio De Giorgi",
Universit\`{a} del Salento and INFN-Lecce, Via Arnesano, 73100 Lecce, 
Italy.}
\ead{paul.h.frampton@gmail.com}
\vspace{10pt}
\begin{indented}
\item[]April 2020
\end{indented}

\begin{abstract}
We discuss a minimal and simple extension of the standard
electroweak model and discuss its uniqueness properties.
Its most readily testable prediction is the existence of
$|Q|= |L|= 2$ gauge bosons, bileptons, which decay into
like-sign lepton pairs. We discuss how a discovery
could be made at the LHC.
\end{abstract}

%
%
%
%
%

\noindent
\section{Introduction}

\bigskip

\noindent
Because the two most popular theoretical models aiming beyond the
standard model - electroweak supersymmetry and large extra dimensions -
have received no encouragement from LHC data, in this talk we shall
discuss what seems to be the most likely first BSM particle.
It is now being searched for by the ATLAS Collaboration.

\bigskip

\noindent
The Bilepton Model was invented as an example of a class of models which turned out, up to variants,
to have one distinct member.

\bigskip

\noindent
\underline{Speculation}

\noindent
Probability LHC will find a new particle : 2/3 likely.
\noindent
If so, guess that the bilepton is 90\% likely so
overall the bilepton can be $(2/3) \times 90\%$ which
means $60\%$ probable.

\bigskip

\noindent
We shall not have time to explain how this model was invented historically but
there is no Royal Road to model building. 
One generally aims for \\
(i)  motivation usually by addressing a question unanswered within the Standard Model.\\
(ii) testability by explicit predictions.\\
Both (i) and (ii) are satisfied by the Bilepton Model.

\noindent
\section{ Bilepton Model}

\noindent
The gauge group is:
\begin{equation}
SU(3)_C \times SU(3)_L \times U(1)_X    \nonumber
\end{equation}

\bigskip

\noindent
The simplest choice for the electric charge is

\begin{equation}
Q = \frac{1}{2} \lambda_L^3 + \left( \frac{\sqrt{3}}{2} \right) 
\lambda_L^8 + X\left(\frac{\sqrt{3}}{\sqrt{2}} \right) \lambda^9    \nonumber
\end{equation}

\bigskip
\noindent
where

\begin{equation}
Tr(\lambda_L^a \lambda_L^b) = 2 \delta^{ab}   \nonumber
\end{equation}

\bigskip

\noindent
and

\begin{equation}
\lambda^9 = \left( \frac{\sqrt{2}}{\sqrt {3}} \right ) ~~ {\rm diag} (1, 1, 1) \nonumber
\end{equation}

\bigskip

\noindent
Thus a triplet has charges $(X+1, X, X-1)$.

\bigskip

\noindent
Leptons are treated democratically in each 
of the three families. They are colour singlets in antitriplets of $SU(3)_L$ :

\bigskip

$(  e^+, \nu_e, e^- )_L$

\bigskip

$( \mu^+, \nu_{\mu}, \mu^- )_L$

\bigskip

$( \tau^+, \nu_{\tau}, \tau^- )_L$

\bigskip

\noindent
All have $X=0$.

\bigskip

\noindent
Quarks in the first family are colour triplets and left-handed triplets plus three singlets

\bigskip

$( u^{\alpha}, d^{\alpha}, D^{\alpha} )_L~~~ (\bar{u}_{\alpha})_L,  (\bar{d}_{\alpha})_L, 
(\bar{D}_{\alpha})_L$

\bigskip

\noindent
Similarly for the second family

\bigskip

$( c^{\alpha}, s^{\alpha}, S^{\alpha} )_L~~~ (\bar{c}_{\alpha})_L,  (\bar{s}_{\alpha})_L, 
(\bar{S}_{\alpha})_L$

\bigskip

\noindent
The X values are for the triplets are $X=-1/3$ and for the singlets
$X=-2/3, +1/3, +4/3$ respectively. The electric charge of the new
quarks $D, S$  is $-4/3$.

\bigskip

\noindent
The quarks of the third family are treated differently. The color triplet quarks
are here in a left-handed antitriplet and three singlets under $SU(3)_L$

\bigskip
\bigskip

$( b^{\alpha}, t^{\alpha}, T^{\alpha} )_L~~~ (\bar{b}_{\alpha})_L,  (\bar{t}_{\alpha})_L, 
(\bar{T}_{\alpha})_L$

\bigskip

\noindent
The antitriplet has $X=+2/3$ and the singlets carry $X=+1/3, -2/3, -5/3$ respectively. 
The new quark $T$ has $Q=5/3$.

\bigskip

\noindent
Before discussing the symmetry breaking to $SU(2)_L \times U(1)_Y$
and the resulting mass 
spectrum, we must explain the nontrivial 
anomaly cancellation of this model which can explain the occurrence
of three quark-lepton families.

\section{ Cancellation of Triangle Anomalies}

\noindent
There are six triangle anomalies which are potentially troublesome; 
in a self-explanatory notation these are diophantine equations\cite{F}

\noindent
$(3_C)^3, (3_C)^2 X, (3_L)^3, (3L)^2 X, X^3, X$ . 

\bigskip

\noindent
The QCD anomaly $(3_C)^3$ is absent because QCD is, as usual, vectorlike. 
$(3_C)^2 X$ vanishes because the quarks are in nine color triplets with net $X =0$
and nine antitriplets also with net $X=0$. The pure $(3_L)^3$ anomaly vanishes 
because there is an equal number of $3_L$ and $3_L^*$.
$(3_L)^2 X$ cancels because the leptons have X =0 and the quarks are in six triplets 
$3_L$ with $X=-\frac{1}{3}$ and three antitriplets $3_L^*$ with $X =+ \frac{2}{3}$. 

\bigskip

\noindent
The $X^3$ cancellation can be checked by a little algebra: 
the three quark families contribute, respectively, $+ 6+ 6 -12 =0$.

\bigskip

\noindent
It is especially interesting that this anomaly cancellation takes place between families. 
Each individual family possesses nonvanishing \\
$(3_L)^3$, $(3L)^2 X$,  $X^3$ anomalies.

\bigskip

\noindent 
Only with the number of families a multiple of three does the overall anomaly vanish.
The imposition of the experimental requirement of  asymptotic freedom of QCD 
then dictates that the number of families be exactly three.

\bigskip

\noindent
The anomaly equations provide a system of six simulataneous
diophantine equations. Algebraic number theory suggests that
such a system almost never has any solution.

\bigskip

\noindent
The fact that one unique solution exists, that underlying the
Bilepton Model. This is the main reason for optimism that this
arrangment of gauge group, embedding of the SM, and
assignment of chiral fermions to irreps may be the one chosen by
Nature.

\bigskip

\noindent
As Einstein said when queried about the 1919 light-bending result
and what would he have thought if the experiments had disagreed
with his general relativity theory: {\it Then I would feel sorry for the Lord. The
theory is correct}. In our experience, such infinite self-belief exists
in some of the best particle theorists.

\bigskip

\section{Upper Limit on Bilepton Mass}

\noindent
The symmetry breaking to the standard model is achieved by a
Vacuum Expectation Value (VEV) of an $X=+1$ triplet
$<\Phi^a> = U\delta^{a3}$. This gives mass $\Lambda_{D,S,T}U$
to the new quarks $D, S, T$ where $\Lambda_i$ are the
Yukawa couplings. It also provides mass to five gauge bosons:
the bileptons $(Y^{\pm\pm}, Y^{\pm})$ and $Z^{'}$.

\bigskip

\noindent
Electroweak breaking is achieved by VEVs
of two triplets $<\phi^a> = v \delta^{a2}$ (with $X=0$)
and $<\phi^{'a}> = v^{'} \delta^{a1}$ (with $X=-1$) and a doublet 
VEV in a sextet with $X=0$\\ 
$<H^{\alpha\beta}> = y\sqrt{10}(\delta^{\alpha1}\delta^{\beta3} 
+ \delta^{\alpha3}\delta^{\beta1})$

\bigskip

\noindent
We note that because of global $L$ symmetry, the $W^+$
and $Y^+$ do not mix. For the same reason, the new quarks
with exotic charges ($D, S, T$) have lepton numbers 
($+2, +2, -2$) respectively.

\bigskip

\noindent
Let the scale of breaking 
$331\rightarrow 321$ be $\mu$. To avoid imaginary
coupling constants with 
$g_i^2 < 0$ which violate unitarity
it is necessary to impose an upper limit on $\mu$ such that

\begin{equation}
\sin^2 \theta (\mu) \leq \frac{1}{4} \nonumber
\end{equation}

\noindent
while at the $Z$ pole the value is

\begin{equation}
\sin^2 \theta(M_Z) \simeq 0.231 \nonumber
\end{equation}

\noindent
which increases using the renormalisation group 
to $\frac{1}{4}$ at $\mu \simeq 4 TeV$. Adopting
this leads to 
\begin{equation}
M_{Y^{\pm\pm}} \leq 2 TeV  \nonumber
\end{equation}

\noindent
by analogy with the electroweak theory
where $M(W^{\pm}) =  80$ GeV which is less
than the $SU(2) \times U(1) \rightarrow U(1)$
breaking scale which is $\sim  248$ GeV. 

\bigskip

\noindent
The estimated upper limit on bilepton mass of $2$ TeV is 
good news for the ongoing LHC bilepton search.

\bigskip

\section{Lower Limit on Bilepton Mass}

\bigskip

\noindent
Perhaps surprisingly the lower limit comes not from 
colliders but from two table-top experiments.

\bigskip

\noindent
Concidentally both experiments have been done
at PSI (= Paul Scherrer Institute). A second coincidence 
is they both give closely the same result for the bilepton
lower mass bound.

\bigskip

\noindent
Firstly there is $\mu^+e^- \rightarrow \mu^-e^+$ which can be
mediated by doubly-charged bilepton exchange. Called
muonium-antimuonium conversion it provides $m_{Y^{\pm\pm}} > 800 GeV$.

\bigskip

\noindent
Secondly there is $\mu^-\rightarrow e^-\nu_e\bar{\nu}_{\mu}$
mediated by singly-charged bilepton exchange
which by Fierz rearrangement is a $(V+A)$
contribution to $\mu^-\rightarrow e^-\bar{\mu}_e\nu_{\mu}$
whose Michel parameter $\xi$ in $(V-\xi A)$ is $1 \geq \xi > 0.997$.
This requires that $m_{Y^{\pm}} > 800 GeV$.

\section{Bilepton Phenomenology at the LHC}

\noindent
A study of bilepton pair production and two or more jets
at the LHC with $\sqrt{s} = 13$ TeV was made, using the Feynman rules
for the Bilepton Model.

\bigskip

\noindent
About 3000 tree-level Feynman graphs were
implemented by \underline{SARAH 4.9.3.}  \\ 
Amplitudes were computed numerically
by \underline{MadGraph}. \\
Simulation of parton showers and hadronisation was made by \underline{HERWIG}.

\bigskip

\noindent
For more details, see\cite{CCCF,CCCF2}

\bigskip

\section{Resonance Bumps}

\bigskip

\noindent
One method \cite{CF} of estimating the numbers of events at ATLAS is to use
old ATLAS data which were analysed to search for a
SSM(=Sequential Standard Model) $Z^{'}$ which was not discovered.
There is some similarity between production of Y and $Z^{'}$. The biggest
difference is that Y must be pair produced so we approximate by using
$\sigma_{SSM}^{Z^{'}} (M(Z^{'} =2M(Y))$. We need to estimate the
brancing ratio (BR) for $(Y \rightarrow e^+e^+, \mu^+\mu^+, \tau^+\tau^+)$.
Because there exist non-leptonic decays $B\rightarrow Q\bar{q}, q\bar{Q}$
where Q is an exotic quark, the BRs depend on the mass M(Q)
of the exotic quarks.

\bigskip

\noindent
We have calculated the branching ratios into each flavour of
like-sign lepton pairs, assuming 800 GeV as the common mass
of the three exotic quarks, see Table 1. If some of the exotic quarks
are more massive than 800 GeV these BRs would be slightly
larger so the estimates in Table 2 will be conservative lower limits
on the event rates.
\begin{table}[h!]
\caption{Branching ratios into like-sign lepton pairs
for each of the three flavours of charged lepton $e, \mu, \tau$.}

\bigskip

\begin{center}

\begin{tabular}{||c|c||}
\hline
M(Y) & BR=branching ratio   \\
GeV & into like-sign lepton pair \\
   & (per flavour) \\
\hline
\hline
800 & 0.33  \\
\hline
1100  &  0.31 \\
\hline
1400 & 0.28 \\
\hline
1700 & 0.25 \\
\hline
2000 & 0.21 \\
\hline
\hline
\end{tabular}
\label{BRS1}

\end{center}

\vspace{0.5in}

\noindent
We re-calculated the numbers of signal events and 
SM background events, using the values of BR from Table 1. 
The results which are our most reliable predictions for the LHC
are displayed in Table 2.

\bigskip

\caption{Numbers of signal and background events at resonance in like-sign
lepton pairs for integrated
luminosity 150/fm. This Table gives our most reliable predictions.}

\bigskip

\begin{center}
\begin{tabular}{||c|c|c|c|c||}
\hline
M(Y) & $\sigma^{Z'}_{SSM}(M_{Z'}=2M(Y))$ & $(BR)^2$ & Signal & Background  \\
GeV & (fb)& & events &  events. \\
\hline
\hline
800 & 100  & 0.109 &1635 & $< 0.01$  \\
\hline
1100  & 20 & 0.096 & 288 & $<0.01$  \\
\hline
1400 &  6 & 0.078 & 70 & $< 0.01$ \\
\hline
1700 & 1 & 0.062 & 9 & $<0.01$ \\
\hline
2000 & 0.6  & 0.044 & 4 & $<0.01$ \\
\hline
\hline
\end{tabular}
\end{center}
\label{BRs}
\end{table}

\bigskip

.

\newpage

\noindent
\underline{{\bf Summary}}

\bigskip
\bigskip

\noindent
Parity violation in weak interactions, chiral fermions
and triangle anomalies underly the Standard Model and
its extension to the Bilepton Model.

\bigskip

\noindent
A possible search for ATLAS and CMS is for 
$Y^{\pm\pm}$ which underly an explanation
of three families in the Bilepton Model 
predicts doubly-charged siblings
$Y^{\pm\pm}$ to accompany $W^{\pm}$.

\bigskip

\noindent
There is the expected mass range

\bigskip

\noindent
$800GeV \leq M_{Y^{\pm\pm}} \leq 2000 GeV$

\bigskip

\noindent 
which renders this new particle accessible to the LHC.

\bigskip

\noindent
What is encouraging about the Bilepton Model
at the LHC is that our calculations of the new signal and
the Standard Model background suggest that obtaining
a $5\sigma$ statistically significant signal for the resonance
seems, at least to a theorist, to be straightforward.

\bigskip

\noindent
As may be recalled from the 2012 announcement by CERN
about discovery of the Higgs Boson, $5\sigma$ is their official requirement
for announcing the discovery of a new particle, as appears readily possible for the
bilepton. Discovery of the bilepton 
would be more transformative than
the Higgs Boson since the latter is already contained
in the Standard Model.  One well-known precedent of a particle discovery was
the $\Omega^-$ particle discovered at Brookhaven
National Laboratory in 1964 which confirmed
Gell-Mann's Eightfold Way theory  proposed
in 1961. That experimental confirmation thus took only three years.
If it were discovered in 2020, the bilepton would have taken 28 years
from 1992 but recall, however, that discovery of the Higgs boson took 
much longer.

\end{document}